\documentclass[a4paper,11pt]{article}
\usepackage{pos}
\newcommand{\MSbar}{{\overline{\rm MS}}}

\newcommand{\be}{\begin{equation}}
\newcommand{\ee}{\end{equation}}
\newcommand{\bea}{\begin{eqnarray}}
\newcommand{\eea}{\end{eqnarray}}
\def\lsim{\mathrel{\rlap{\lower4pt\hbox{\hskip1pt$\sim$}}
    \raise1pt\hbox{$<$}}}                
\def\slashed{{/}\mskip-10.0mu}
\def\qcircslash{\slashed {q\mskip -5mu ^{^\circ}}}
\def\qcirc{ {q\mskip -5mu ^{^\circ}}}
\def\archcoth{\rm arccoth}

\title{Mass effects on the QCD $\beta$-function}

\author*[a, b]{M.~Costa}
\author[a]{D.~Gavriel}
\author[a]{H.~Panagopoulos}
\author[a]{G.~Spanoudes}

\affiliation[a]{
Department of Physics, University of Cyprus,\\
1 Panepistimiou Avenue, Aglantzia, P. O. Box 2109, CY-1678 Nicosia, Cyprus}

\affiliation[b]{Department of Chemical Engineering, Cyprus University of Technology,\\
30 Archbishop Kyprianou Str., CY-3036, Limassol, Cyprus}

\emailAdd{kosta.marios@ucy.ac.cy}
\emailAdd{gavriel.demetrianos@ucy.ac.cy}
\emailAdd{panagopoulos.haris@ucy.ac.cy}
\emailAdd{spanoudes.gregoris@ucy.ac.cy}

\abstract{In this study we present lattice results on the QCD $\beta$-function in the presence of quark masses. The $\beta$-function is calculated to three loops in perturbation theory and for improved lattice actions; it is extracted from the renormalization of the coupling constant $Z_g$. The background field method is used to compute $Z_g$, where it is simply related to the background gluon field renormalization constant $Z_A$. We focus on the quark mass effects in the background gluon propagator; the dependence of the QCD $\beta$-function on the number of colors $N_c$, the number of fermionic flavors $N_f$ and the quark masses, is shown explicitly. The perturbative results of the QCD $\beta$-function will be applied to the precise determination of the strong coupling constant, calculated by Monte Carlo simulations removing the mass effects from the nonperturbative Green's functions.
\begin{center}
\includegraphics[scale=0.45]{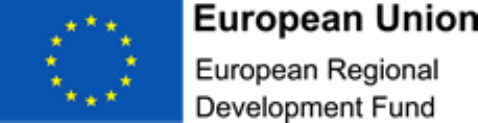}
\includegraphics[scale=0.45]{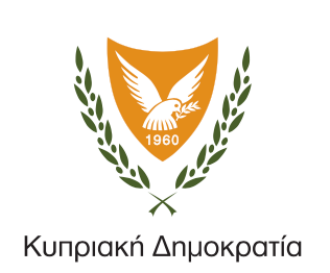}
\includegraphics[scale=0.45]{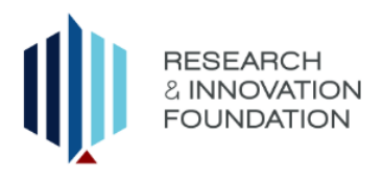}
\includegraphics[scale=0.45]{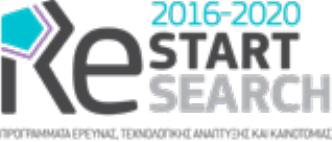}
\end{center}
}

\FullConference{The 40th International Symposium on Lattice Field Theory (Lattice 2023)\\
July 31st - August 4th, 2023\\
Fermi National Accelerator Laboratory\\}

\begin{document}
\maketitle

\section{Introduction -- Motivation}

The renormalized $\beta$-function, describing the dependence of the renormalized coupling constant $g^R$ on the scale of the renormalization scheme~\cite{Luscher:1995np}, plays a crucial role in understanding the underlying dynamics of QCD across different momentum regimes. It encodes the underlying dynamics of QCD from low to high momentum regions. Nonperturbative estimations of the strong coupling $\alpha$, in several renormalization schemes, through numerical simulations of the corresponding lattice theories, are being studied by a number of groups at present (see e.g., Refs.~\cite{DallaBrida:2020pag, DelDebbio:2021ryq} and references therein).

The three-loop bare QCD $\beta$-function~\cite{Bode:2001at} can be extracted from the two-loop calculation of the renormalization factor $Z_g$, which relates the bare running coupling $\alpha_0$ to the $\MSbar$-renormalized running coupling $\alpha_{\MSbar}$ ($\bar \mu$ is the $\MSbar$ renormalization scale and $a$ is the lattice spacing) through: 
     \begin{equation}
         \alpha_0= Z_g^2 (g_0, a \bar{\mu})\times \alpha_{\MSbar} 
     \end{equation} 
     
The inclusion of quark masses makes this calculation even more complicated~\cite{Dietrich:2009ns}. Note that we are interested in the discretization errors proportional to the quark mass ($O(a m)$ effects) on the $\beta$-function. For simplicity of notation, we denote all flavor masses by $m$; the case of different flavor masses can be trivially recovered from our results. This new direction is very important due to the fact that the $O(a m)$ effects will be removed from the nonperturbative Green’s functions entering the strong coupling, allowing for a more precise determination. Furthermore, removing  $O(a m)$  effects will improve importantly any quantity that is calculated using Wilson-type fermions~\cite{DallaBrida:2022eua}. 

\section{Computational setup and methods used to calculate $Z_g$}

The renormalized $\beta$-function and the bare $\beta$-function on the lattice ($\beta_L(g_0)$) are defined:
\begin{equation}
 \beta(g_{\rm \overline{MS}}) = \bar{\mu} \frac{dg_{\rm \overline{MS}}}{d\bar{\mu}} \Big|_{a,g_0}, \qquad
\beta_L(g_0)= -a \frac{dg_0}{da} \Big|_{\bar{\mu}, g_{\rm \overline{MS}} } 
\label{beta}
\end{equation}
In the asymptotic limit, one can write the expansion of Eq.~(\ref{beta}) in powers of $g_0$:
\begin{eqnarray}
\beta_L(g_0) &&=\,-b_0 \,g^3_0 -b_1 \,g_0^5 - b_2^{L}\,g_0^7 - ..., \\
\beta(g_{\rm \overline{MS}}) &&=\,  -b_0 \,g_{\rm \overline{MS}}^3 -b_1 \,g_{\rm \overline{MS}}^5 - b_2\,g_{\rm \overline{MS}}^7 + ... 
\end{eqnarray}
The coefficients $b_0, b_1$ are well-known universal constants (regularization independent) for the massless case; $b_i^{L}$ ($i \ge 2 $) (regularization dependent) must be calculated perturbatively. $\displaystyle \beta_L(g_0)$ and $\displaystyle \beta(g_{\rm \overline{MS}})$ can be related using the renormalization function  $Z_g$, that is:
\begin{equation}
\beta_{L}(g_0) = \left( 1 - g_0^2 \,\,{\partial \ln Z_g^2 \over \partial g_0^2} \right)^{-1}Z_g \,\,\beta(Z_g^{-1} g_0)
\label{beta1}
\end{equation}
The most convenient and economical way to proceed with the calculation of $Z_g(g_0,a\bar{\mu})$ 
is to use the Background Field ($BF$) technique~\cite{Ellis:1983af, Luscher:1995vs}, in which the following relation is valid. 
\begin{equation}
Z_A(g_0,a\bar{\mu})   Z_g^2(g_0, a\bar{\mu}) = 1 
\end{equation}
where $Z_A$ is the $BF$ renormalization function. In the lattice version of the $BF$ technique, the link variable takes the form: $U_{x, x+\mu}= e^{i a g_0 Q_{\mu}(x)}\cdot e^{i a A_{\mu}(x)}\,$ ($Q_{\mu}$: quantum field, $A_{\mu}$: background field). In this framework, instead of calculating $Z_g$, it suffices to compute $Z_A$. Note that the inclusion of quark masses adds an additional layer of complexity to this calculation, and we are particularly interested in understanding $O(a m)$ effects on the $\beta$-function.

Therefore, our attention is directed towards the 2-point $BF$ 1PI Green's function, denoted as $\langle A_\mu(x) A_\nu(y) \rangle$ (in a slight abuse of notation, cf.~\cite{Luscher:1995vs}), where we focus on the quark mass effects of $O (a m)$. Currently, we have completed the computation of the one-loop quantum correction for this Green's function, and the two-loop calculations are in progress. 

Our computations are carried out within the lattice regularization, utilizing the clover improved action for fermions and a class of Symanzik improved gauge actions. The clover action reads, in standard notation:
\bea
S_F &=& \sum_{f}\sum_{x} (4r+m_f)\bar{\psi}_{f}(x)\psi_f(x)\nonumber \\
&-& {1\over 2}\sum_{f}\sum_{x,\,\mu}\bigg{[}\bar{\psi}_{f}(x) \left( r - \gamma_\mu\right)
U_{x,\, x+\mu}\psi_f(x+{\mu}) 
+\bar{\psi}_f(x+{\mu})\left( r + \gamma_\mu\right)U_{x+\mu,\,x}\psi_{f}(x)\bigg{]}\nonumber \\
&-& {1\over 4}\,c_{\rm SW}\,\sum_{f}\sum_{x,\,\mu,\,\nu} \bar{\psi}_{f}(x)
\sigma_{\mu\nu} {\hat F}_{\mu\nu}(x) \psi_f(x),
\label{clover}
\eea
The Wilson parameter $r$ is set to $r=1$; $f$ is a flavor
index; $\sigma_{\mu\nu} =[\gamma_\mu,\,\gamma_\nu]/2$\,; the clover
coefficient $c_{\rm SW}$ is kept as a free parameter throughout. Powers of
the lattice spacing $a$ have been omitted and may be directly
reinserted by dimensional counting. The tensor $\hat{F}_{\mu\nu}$
is proportional to a lattice representation of the gluon field tensor; it is defined through: ${\hat F}_{\mu\nu} \equiv {1\over{8}}\,(Q_{\mu\nu} - Q_{\nu\mu})$, where $Q_{\mu\nu}$ is the sum of the plaquette loops:
\begin{eqnarray}
Q_{\mu\nu} &=& U_{x,\, x+\mu}U_{x+\mu,\, x+\mu+\nu}U_{x+\mu+\nu,\, x+\nu}U_{x+\nu,\, x} \nonumber \\
&+& U_{ x,\, x+ \nu}U_{ x+ \nu,\, x+ \nu- \mu}U_{ x+ \nu- \mu,\, x- \mu}U_{ x- \mu,\, x} \nonumber \\
&+& U_{ x,\, x- \mu}U_{ x- \mu,\, x- \mu- \nu}U_{ x- \mu- \nu,\, x- \nu}U_{ x- \nu,\, x} \nonumber \\
&+& U_{ x,\, x- \nu}U_{ x- \nu,\, x- \nu+ \mu}U_{ x- \nu+ \mu,\, x+ \mu}U_{ x+ \mu,\, x}
\end{eqnarray}

For the gauge fields we employ the Symanzik improved action, involving Wilson loops with 4 and 6 links ($1 \times 1$ plaquettes and $1 \times 2$ rectangles, respectively), which is given by the relation
\begin{eqnarray}
\hspace{-1cm}
S_G&=&\frac{2}{g_0^2} \Bigl[ c_0 \sum_{\rm plaquette} {\rm Re\,Tr\,}\{1-U_{\rm plaquette}\} 
+ c_1 \sum_{\rm rectangle} {\rm Re \, Tr\,}\{1- U_{\rm rectangle}\} \Bigr]
\label{Symanzik}
\end{eqnarray}
The coefficients $c_0$ and $c_1$ can in principle be chosen arbitrarily, subject to the following normalization condition, which ensures the correct classical continuum limit of the action:
\begin{equation}
c_0 + 8 c_1  = 1.
\label{norm}
\end{equation}
Particular choices of values for $\{c_0$, $c_1\}$ are employed in our calculaltions ($\{1,0\}$: Wilson gluons, $\{5/3 , -1/12\}$: Symanzik tree-level improved action and $\{3.648, -0.331\}$: Iwasaki action) .
\section{One-loop Results}

\begin{figure}[ht!]
\centering
        \includegraphics[scale=0.70]{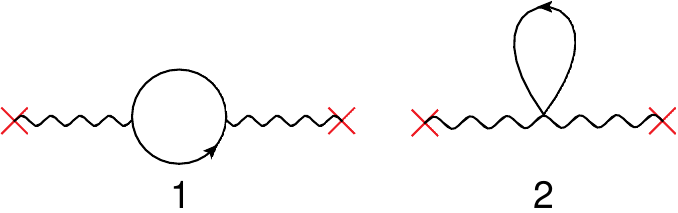}
\caption{
One-loop Feynman diagrams for fermion contributions to $\langle A_\mu A_\nu \rangle$.  
A solid line represents quarks. Wavy lines ending on a cross represent background gluons. 
Each diagram is meant to be symmetrized over the color indices, Lorentz indices and momenta of the two external background fields.
}
\label{figoneloop}
\end{figure}

The mass effects, which contribute to the 2-point Green's function $\langle A_\mu(x)  A_\nu(y) \rangle$, are associated with the Feynman diagrams with at least one fermion line. At one-loop order, fermion contributions to  $\langle A_\mu(x)  A_\nu(y) \rangle^{1-loop}$ come from the sum of the Feynman diagrams in Figure ~\ref{figoneloop}. 

The one-loop result of the fermion contributions to 2-pt lattice Green's function is: 
\begin{equation}
\langle A^\alpha_\mu A^\beta_\nu \rangle^{1-loop} =\delta^{\alpha \beta} N_f \left(\delta_{\mu \nu} q^2  -q_\nu q_\mu \right) \Big\{F_1(a q) + F_2\left(\frac{m^2}{q^2}\right) + am \Big[F_3 (a q) + F_4 \left(\frac{m^2}{q^2}\right) \Big]
\Big\},
\label{oneloopGF}
\end{equation}
where:
\begin{eqnarray}
F_1(a q) &=& -0.0137322+0.0050467\,c_{sw}-0.0298435\,c_{sw}^2+\frac{2}{3}\frac{1}{16\pi^2}\log(a^2 q^2)\nonumber\\\nonumber
F_2\left(\frac{m^2}{q^2}\right) &=& \frac{8}{3}\frac{1}{16\pi^2} \frac{m^2}{q^2} - \frac{8}{3}\frac{1}{16\pi^2} \left(-\frac{1}{2} +\frac{m^2}{q^2} \right) \sqrt{1 + 4\frac{m^2}{q^2}} \, \archcoth\left(\sqrt{1 + 4\frac{m^2}{q^2}}\right) +\frac{2}{3}\frac{1}{16\pi^2}\log\left(\frac{m^2}{q^2}\right)\nonumber\\\nonumber F_3(a q) &=& 0.0272837-0.0223503 c_{sw} + 0.0070667 c_{sw}^2 - (1 -  c_{sw})\frac{2}{16\pi^2}\log\left(a^2 q^2\right)\nonumber\\\nonumber
F_4 \left(\frac{m^2}{q^2}\right) &=& -\frac{4}{16 \pi^2} \frac{m^2}{q^2} +\frac{4}{16 \pi^2}  \left[(-1 + c_{sw}) \left(1 + 4\frac{m^2}{q^2}\right)
+ 4\left(\frac{m^2}{q^2}\right)^2 \right] \frac{\archcoth\left(\sqrt{1 + 4\frac{m^2}{q^2}}\right)}{\sqrt{1 + 4\frac{m^2}{q^2}}} \nonumber\\\nonumber
&&\hspace{3.5cm}- (1 -  c_{sw})\frac{2}{16\pi^2}\log\left(\frac{m^2}{q^2}\right)
\end{eqnarray}
Since Eq.~(\ref{oneloopGF}) stemming from diagrams of closed fermion loops, the one-loop results are independent of the Symanzik coefficients. However, our two-loop calculations employed values for the Symanzik coefficients which are commonly used.

We define the $BF$ coupling to one-loop order as (for $c_{sw}=1+O(g_0^2)$): 
\begin{equation}
g_{BF}^2(q,m) = g_0^2 + g_0^4 \Big\{F_1(a q) + F_2\left(\frac{m^2}{q^2}\right) + a m \Big[F_3(a q) + F_4 \left(\frac{m^2}{q^2}\right)  \Big]  \Big\} \Big|_{c_{sw}=1} + O(g_0^6)
\end{equation}
$g_{BF}^2(q,m)$ can be expressed in terms of the renormalized coupling $g_\MSbar$ through $Z_g^{L,\MSbar}$ where $g_\MSbar^2 = Z_g^{L,\MSbar} g_0^2, \quad Z_g^{L,\MSbar} = 1 - g_0^2 \left(b \log\left(a^2 \bar \mu^2\right) - a \, m \, b_g \right)  + O(g_0^4) $. It easy to show that in order to remove the unwanted lattice contributions ($\log(a)$ and $(a m)$) the coefficients $b$ and $b_g$ must be~\cite{Luscher:1996sc}:
\begin{equation}
b = -\frac{1}{24 \pi^2}, \quad b_g = 0.01200
\end{equation}

At one loop order, expressing $g_{BF}^2(q,m)$ in terms of renormalized quantities ($g_\MSbar$, $m_\MSbar = m \left(1 - \frac{1}{2} a m \right)$ and taking the limit $z \to \infty$ $\,(z ={m_\MSbar^2}/{q^2})$ we get: 
\begin{equation}
\lim_{z \to \infty} g_{BF}^2(q,m_\MSbar) = g_\MSbar^2  + g_\MSbar^4 \left(-0.0314928 + \frac{1}{24\pi^2}\log\left( m_\MSbar^2/ \bar \mu^2 \right) \right)  + O(a^2,g_\MSbar^6)
\label{MassD}
\end{equation}
Equation~(\ref{MassD}) distinctly illustrates the logarithmic mass dependence exhibited by heavy quarks as we approach the continuum limit.

The one-loop results provide valuable insights into the quark mass effects on the QCD $\beta$-function; the logarithmic mass dependence in the $BF$ coupling shows the significance of heavy quark behavior in the continuum limit. While the one-loop results lay a foundational understanding, ongoing two-loop calculations are crucial for refining the precision of the QCD $\beta$-function and completing the picture of quark mass effects on the renormalization of the coupling constant within the lattice QCD framework.

\section{Two-loop Calculations}

\begin{figure}[ht!]
\centering
\includegraphics[scale=1.00]{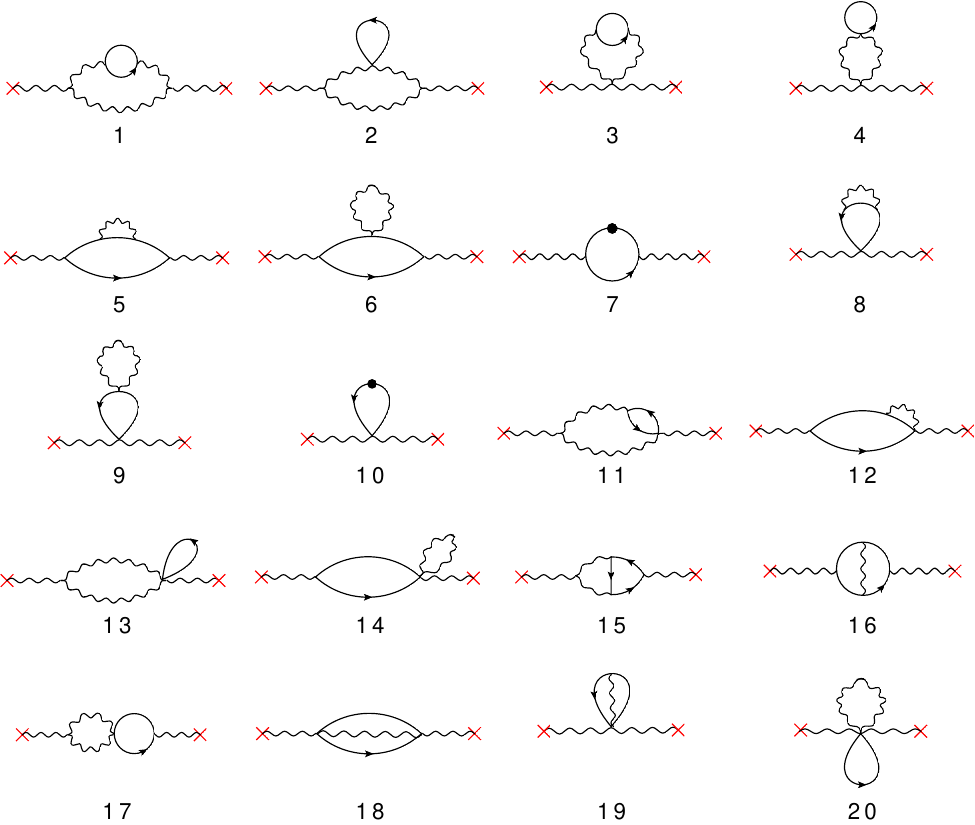}
\caption{
Two-loop Feynman diagrams for the fermion contributions to  $\langle A_\mu A_\nu \rangle$.
A wavy (solid) line represents gluons (quarks). Wavy lines ending on a cross represent background gluons. A solid circle is the one-loop fermion mass counterterm.
Each diagram is meant to be symmetrized over the color indices, Lorentz indices, and momenta of the two external background fields.
}
\label{figtwoloop}
\end{figure}

The computation of the two-loop Feynman diagrams is currently in progress. The result of the fermion contribution to the two-loop 2-point lattice Green's function is obtained as the sum of twenty Feynman diagrams, as shown in Figure ~\ref{figtwoloop}.

Since we are interested in the $O(a m)$ corrections, we use the relation for the tree-level fermion propagator in momentum space: 
\begin{equation}
\langle \psi  \bar\psi \rangle = \frac{-i \,\qcircslash + M(q,m)}{\qcirc^2 + M(q,m)^2}, 
\label{propF}
\end{equation}
where: $\qcircslash = \,\sum_\mu\gamma_\mu  \,\frac{1}{a}\sin(a q_\mu)$ and $M(q,m) = m + \frac{2}{a} \sum_\mu \sin^2(a q_\mu/2)$.

To study the mass effects, we expand the denominator of Eq.~(\ref{propF}) with respect to $m$ and we obtain Eq.~(\ref{propFm}).
\begin{eqnarray}  
\frac{1}{\qcirc^2 + M(q,m)^2} &=& \frac{1}{\qcirc^2 + M(q,0)^2} \left(1 - \frac{4 m \frac{1}{a} \sum_\mu \sin^2(a q_\mu/2) }{\qcirc^2 + M(q,0)^2} + O(a^2 m^2)  \right) 
\label{propFm}
\end{eqnarray}

One main difficulty in this computation, as compared to the $O((am)^0)$ calculation, stems from the fact that the fermion propagator now contains contributions of $O(q^{-2})$; this amplifies the presence of potential IR divergences, which must be carefully addressed. Also, the sheer number of terms which must be integrated over the two loop momenta is of the order of $\sim 10^6$; this has necessitated the creation of special-purpose integration routines, in order to overcome the severe contraints on CPU and memory. These ongoing two-loop calculations are anticipated to offer a more thorough understanding of the quark mass effects in the renormalization of the coupling constant. However, it is essential to note also that the computational challenges are further amplified by the distinct methodologies and manipulations required for each diagram. Particularly, the "diamond" diagrams (diagrams 15 and 16 in Figure ~\ref{figtwoloop}) stand out as the most challenging within this computation.

\section{Conclusions}

In this study, we examined the QCD $\beta$-function on the lattice, taking into account quark masses and emphasizing $O(a m)$ effects. By employing the Background Field method, we derived the $\beta$-function from the renormalization of the coupling constant $Z_g$. Our one-loop results for the 2-point lattice Green's function show the importance of the quark mass effects in the $\beta$-function. The ongoing two-loop calculations play a crucial role in refining the precision of the QCD $\beta$-function and advancing our comprehensive understanding of quark mass effects. We anticipate that our perturbative results will contribute to the precise determination of the strong coupling constant in numerical simulations, thereby enhancing the accuracy of nonperturbative Green's function calculations in lattice QCD. 
\\\\
{\bf Acknowledgements:} 
 This work is funded by the European Regional Development Fund and the Republic of Cyprus through the Research and Innovation Foundation (Project:  EXCELLENCE/0421/0025). We thank Dr. Mattia Dalla Brida for fruitful discussions and helpful comments.

\end{document}